\documentclass[a4paper,12pt]{article}

\usepackage[english]{babel}
\usepackage[T1]{fontenc}
\usepackage{authblk}
\usepackage{natbib}
\usepackage{here}
\usepackage{setspace}
\usepackage{multirow}
\usepackage{lmodern}
\usepackage{amsmath, amssymb}
\usepackage{bm}
\usepackage{pdflscape}
\usepackage{authblk}
\usepackage{color}
\usepackage[table]{xcolor}
\usepackage{tikz}

\usepackage{booktabs}
\usepackage{afterpage}
\usepackage{textcomp}
\usepackage{float}
\usepackage{enumerate}
\usepackage{rotating}

\definecolor{babyblue}{rgb}{0.54, 0.81, 0.94}
\definecolor{babypink}{rgb}{0.96, 0.76, 0.76}

\usepackage[margin=1in]{geometry}

\usepackage{amsmath}
\usepackage{graphicx}
\usepackage[colorlinks=true, allcolors=blue]{hyperref} 

\usepackage{caption}
\usepackage{subcaption}

\definecolor{Gray}{gray}{0.85}
\newcolumntype{a}{>{\columncolor{Gray}}c}
\newcolumntype{d}{>{\columncolor{white}}c}

\begin{document}

    \begin{center}
        \vspace*{1cm}
        \large
	    \textbf{Guidelines for the use of spatially-varying coefficients in species distribution models}\\
         \normalsize
           \vspace{5mm}
	    Jeffrey W. Doser\textsuperscript{1, 2}, Marc K{\'e}ry\textsuperscript{3}, Sarah P. Saunders\textsuperscript{4}, Andrew O. Finley\textsuperscript{2,5,6},  Brooke L. Bateman\textsuperscript{4}, Joanna Grand\textsuperscript{4}, Shannon Reault\textsuperscript{4}, Aaron S. Weed\textsuperscript{7}, Elise F. Zipkin\textsuperscript{1, 2}
         \vspace{5mm}
    \end{center}
    \small
	     \textsuperscript{1}Department of Integrative Biology, Michigan State University, East Lansing, MI, USA \\
         \textsuperscript{2}Ecology, Evolution, and Behavior Program, Michigan State University, East Lansing, MI, USA \\
         \textsuperscript{3}Swiss Ornithological Institute, Sempach, Switzerland \\
         \textsuperscript{4}Science Division, National Audubon Society, New York, NY, USA \\
         \textsuperscript{5}Department of Forestry, Michigan State University, East Lansing, MI, USA \\
         \textsuperscript{6}Department of Statistics and Probability, Michigan State University, East Lansing, MI, USA \\
         \textsuperscript{7}Northeast Temperate Inventory and Monitoring Network, National Park Service, Woodstock, VT, USA \\  
         \noindent \textbf{Corresponding Author}: Jeffrey W. Doser, email: doserjef@msu.edu; ORCID ID: 0000-0002-8950-9895 \\

\section*{Data Availability Statement}

All data and code associated with this manuscript are available on GitHub (\url{https://github.com/doserjef/Doser_et_al_2023_SVC}) and will be posted on Zenodo upon acceptance. 

\section*{Acknowledgements}

This work was supported by National Science Foundation (NSF) grants DBI-1954406, DMS-1916395, and DEB-2213565.

\section*{Biosketch}

Jeffrey W. Doser is a statistical ecologist who develops and applies hierarchical Bayesian models to inform wildlife and natural resource conservation across spatial scales. 
         
\newpage

\section*{Abstract}

\subsection*{Aim}

Species distribution models (SDMs) are increasingly applied across macroscales using detection-nondetection data. Such models typically assume that a single set of regression coefficients can adequately describe species-environment relationships and/or population trends. However, such relationships often show nonlinear and/or spatially-varying patterns that arise from complex interactions with abiotic and biotic processes that operate at different scales. Spatially-varying coefficient (SVC) models can readily account for variability in the effects of environmental covariates. Yet, their use in ecology is relatively scarce due to gaps in understanding the inferential benefits that SVC models can provide compared to simpler frameworks.

\subsection*{Innovation}

Here we demonstrate the inferential benefits of SVC SDMs, with a particular focus on how this approach can be used to generate and test ecological hypotheses regarding the drivers of spatial variability in population trends and species-environment relationships. We illustrate the inferential benefits of SVC SDMs with simulations and two case studies: one that assesses spatially-varying trends of 51 forest bird species in the eastern US over two decades and a second that evaluates spatial variability in the effects of five decades of land cover change on Grasshopper Sparrow (\textit{Ammodramus savannarum}) occurrence across the continental US.

\subsection*{Main conclusions}

We found strong support for SVC SDMs compared to simpler alternatives in both empirical case studies. Factors operating at fine spatial scales, accounted for by the SVCs, were the primary divers of spatial variability in forest bird occurrence trends. Additionally, SVCs revealed complex species-habitat relationships with grassland and cropland area for Grasshopper Sparrow, providing nuanced insights into how future land use change may shape its distribution. These applications display the utility of SVC SDMs to help reveal the environmental factors that drive species distributions across both local and broad scales. We conclude by discussing the potential applications of SVC SDMs in ecology and conservation.

\newpage

\section*{Introduction}

Elucidating the factors that drive species distributions is a fundamental objective of ecology. Species distribution models (SDMs) are the primary tool to study where and why species occur across space and time \citep{guisan2000predictive}. While SDMs can leverage a variety of data types (e.g., presence-only, abundance), they are commonly used with presence-absence (or detection-nondetection) data in a parametric generalized-linear model (GLM)-based framework, allowing for assessments of species-environment relationships and probabilities of local-level occurrence. Parametric SDMs often describe species-environment relationships via a single set of regression coefficients (e.g. linear and quadratic terms) across the spatial extent of the data set, resulting in a set of linear or unimodal response curves along all environmental predictors included in the model \citep{guisan2002generalized}. When the spatial extent encompasses the entirety of the species range, the combination of such species-environment response curves characterizes the multivariate realized environmental niche of a species \citep{guisan2017habitat}. However, when fitting SDMs across large spatial extents, a single set of linear and quadratic regression coefficients may not adequately describe species-environment relationships, which can result from the true relationship taking a complex, nonlinear form, variability in the relationship across space (i.e., the relationship is context dependent, or nonstationary; \citealt{rollinson2021working}), or simply due to inconsistencies in habitat classification across space (e.g. when ``arid grassland'' does not refer to the same exact habitat type across all regions of a modeled domain). 

Complex, nonlinear and/or spatially-varying species-environment relationships can arise from a variety of abiotic and biotic processes that operate at different scales \citep{osborne2002should, miller2012species, clark2020nonlinear}. Interactions with abiotic factors, such as historical disturbance regimes, landscape composition/configuration, fine-scale habitat characteristics (e.g., vegetation quality), and environmental conditions (e.g., soil content, temperature) can result in varying effects of environmental factors on species across their ranges \citep{rollinson2021working}. Given spatial heterogeneity in resource availability, effects of environmental factors on species occurrences may be stronger in areas with limited resources compared to areas where resources are more abundant \citep{pease2022exploring}. If climate shapes species distributions, effects of climate change on species should be strongest in areas near their climatic limits \citep{amburgey2018range}. For example, \cite{sultaire2022spatial} found spatial variation in the effects of increasing temperature and snow cover duration on snowshoe hare (\textit{Lepus americanus}) occurrence, suggesting that climate limits their distribution in different ways across the species range. Alternatively, spatially-varying species-environment relationships may arise from biotic processes such as local genetic adaptations or spatial variation in species interactions (e.g., predation, competition). \cite{pease2022exploring} found that spatial variability in the effect of forest cover on white-tailed deer (\textit{Odocoileus virginianus}) occurrence across North Carolina was partially driven by variation in predation pressure across the state. Failing to account for such spatially-varying species-environment relationships when present can lead to misleading inferences on the abiotic factors that influence where species occur, which could have important implications on management and conservation recommendations \citep{rollinson2021working}.

In addition to characterizing species-environment relationships, monitoring data are often used in SDMs to quantify occurrence trends. Nonlinear and/or spatially-varying occurrence trends primarily arise from spatio-temporal changes in abiotic and biotic factors that influence the species of interest (e.g., differences in the amount of land-use change across a species distribution). However, such patterns can also arise from complex relationships to such abiotic and biotic factors (e.g., the effect of land-use change is different in one part of a species range compared to another). Quantifying spatial variability in occurrence over time is a common objective of biodiversity monitoring programs \citep{bled2013modeling, babcock2016modeling, meehan2019spatial}, as such insights can help generate hypotheses about the drivers of population changes (e.g., \citealt{crossley2021complex}), identify priority areas for conservation or restoration (e.g., \citealt{ethier2017spatiotemporal}), and provide insights into where additional monitoring effort is needed to reduce uncertainty \citep{reich2018integrating}.

Throughout the ecological literature, numerous methods have been used to test hypotheses about spatial variability and nonlinearity in species-environment relationships or occurrence trends. GLMs with interactions between abiotic and/or biotic variables are simple, yet intuitive, ways to assess non-linear and/or spatially-varying relationships \citep{spake2023understanding}. In addition, estimating separate slopes (fixed or random) across pre-defined strata (e.g., ecoregions, management units) is another common alternative (e.g., \citealt{smith2021north}). Further, nonlinear functions (e.g., thresholds) can be readily incorporated into parameteric SDMs \citep{hostetler2015improved}. However, these approaches require \textit{a priori} knowledge of covariates that interact with the variables of interest, the functional forms of such relationships, and/or the spatial resolution of variability in the relationship, most of which are unknown prior to analysis. Further, specifying interactions between multiple drivers operating at different spatial scales is particularly critical when working across macroscales, but difficult to accomplish using the aforementioned approaches. More flexible approaches that can readily accommodate complex species-environment relationships without \textit{a priori} knowledge of all interacting variables and the nature of their interactions are thus needed.

Accordingly, there has been widespread use of machine-learning approaches, such as MaxEnt \citep{phillips2006maximum} and random forests \citep{liaw2002classification}, to model species distributions while accounting for complex species-environment relationships. MaxEnt uses combinations of different ``feature classes'', or mathematical functions, on covariates to characterize nonlinear relationships in presence-only data, and random forests fit ensembles of classification or regression trees within partitions of the data based on covariate space \citep{valavi2021modelling}. Both approaches are widely used in ecology and conservation, but they are limited in their ability to provide uncertainty estimates of species-environment relationships and/or occurrence trends; they require all interacting variables to be known and incorporated into the model; and they do not account for imperfect detection (i.e., the failure to observe a species at a site when it is present; \citealt{kery2011towards, kellner2014accounting}). 

In this paper, we discuss the use of spatially-varying coefficients (SVCs) in SDMs, a highly flexible approach for modeling nonlinear and/or spatially-varying species-environment relationships, regardless of how they arise. SVC models are intuitive extensions of GLMs that allow regression coefficients to vary smoothly across space. By fitting SVC models within a hierarchical Bayesian framework, we can generate predictions of species-environment effects and/or trends across a spatial region of interest with full uncertainty propagation, while simultaneously accounting for widespread observation errors, such as those due to imperfect detection. Recent studies suggest an increasing interest in this flexible framework for a variety of ecological applications due to the prevalence of heterogeneity in species-environment relationships across macroscales (e.g., \citealt{meehan2019spatial, rollinson2021working, sultaire2022spatial}). However, a comprehensive understanding of the inferential benefits SVCs can provide compared to alternative approaches is lacking. \cite{thorson2023spatially} recently highlighted seven ecological questions that can be addressed with spatially-varying coefficients. Here, we build on their work by explicitly demonstrating the inferential benefits of SVC SDMs compared to simpler approaches with simulations and two case studies on breeding birds in the US. We conclude with a discussion of practical guidelines on when to use SVC SDMs instead of, or in addition to, simpler alternatives.

\section*{What are spatially-varying coefficients?}

Our motivating case studies focus on imperfectly observed bird species. As such, we discuss SVCs in the context of occupancy models \citep{mackenzie2002, tyre2003improving}, a specific form of hierarchical GLM that models imperfect detection, although our findings are directly extensible to all parametric SDMs (i.e., GLMs). Here we give a brief overview of SVC occupancy models. See \cite{doser2023SVC} for full statistical details. 

Let $\bm{s}_j$ denote the spatial coordinates of site $j$, where $j = 1, \dots, J$, which are each sampled across $t = 1, \dots, T$ primary time periods (henceforth ``seasons''). Note that data may be obtained for only one season (i.e., $T = 1$), or for multi-season data sets in which sites do not need to be sampled every season (i.e., missing values are allowed). To account for imperfect detection, $k = 1, \dots, K_t(\bm{s}_j)$ sampling replicates are obtained at site $j$ during season $t$ to estimate whether a nondetection of the target species is truly an absence \citep{mackenzie2002, tyre2003improving}. Note that the number of replicates can vary across site/season combinations, and multiple replicates within a season are not necessarily needed at every site (e.g., \citealt{von2023mixed}). Such replicates typically come in the form of multiple visits to a site over a short period of time within a season, but other forms of replication such as spatial sub-sampling and multiple observers are possible \citep{mackenzie2017occupancy}. Let $y_{t, k}(\bm{s}_j)$ denote the observed detection (1) or nondetection (0) of a study species at site $j$ during survey $k$ in season $t$, and let $z_t(\bm{s}_j)$ denote the true presence (1) or absence (0) of the species at site $j$ during season $t$. Note we assume $z_t(\bm{s}_j)$ does not change across replicate surveys within a given season (i.e., the ``closure'' assumption). We model the observed data $y_{t, k}(\bm{s}_j)$ conditional on the true occurrence status of the species at site $j$ during season $t$ ($z_t(\bm{s}_j)$). Specifically, we have
\begin{equation}
  y_{t, k}(\bm{s}_j) \sim \left\{
    \begin{matrix}    
      0,\hfill &z_t(\bm{s}_j) = 0 \\
      \text{Bernoulli}(p_{t, k}(\bm{s}_j)),\hfill &z_t(\bm{s}_j) = 1 \end{matrix} \right.,
\end{equation}

where $p_{t, k}(\bm{s}_j)$ is the probability of detecting the species at site $j$ during replicate survey $k$ in season $t$. We model detection probability as a function of site, season, and/or survey-level (i.e., observation-level) covariates according to 
\begin{equation}\label{pDet}
  \text{logit}(p_{t, k}(\bm{s}_j)) = \bm{v}_{t, k}(\bm{s}_j)\bm{\alpha}, 
\end{equation}

where $\bm{\alpha}$ is a vector of regression coefficients (including an intercept) that describe the effect of covariates $\bm{v}_{t, k}(\bm{s}_j)$ on detection. 

The true occurrence status $z_t(\bm{s}_j)$ is a partially observed variable, such that if $y_{t, k}(\bm{s}_j) = 1$, we know $z_t(\bm{s}_j) = 1$ (since we assume no false positives), but if $y_{t, k}(\bm{s}_j) = 0$ we do not know if the species is truly absent from the site, or if we failed to detect it. Accordingly, we model $z_t(\bm{s}_j)$ as 
\begin{equation}\label{z}
    z_t(\bm{s}_j) \sim \text{Bernoulli}(\psi_t(\bm{s}_j)), 
\end{equation}

where $\psi_t(\bm{s}_j)$ is the occurrence probability of the species at site $j$ during season $t$. When fitting occupancy models, we can estimate species-environment relationships through the effect of covariates on occurrence probability, $\psi_t(\bm{s}_j)$, within a GLM framework. For simplicity, consider a single environmental variable, $\text{x}_t(\bm{s}_j)$, that varies across each spatial location $\bm{s}_j$ and season $t$ (e.g., temperature, precipitation). An SVC occupancy model has the form 

\begin{equation}\label{psi}
   \text{logit}(\psi_t(\bm{s}_j)) = \beta_0 + \eta_t + \text{w}_0(\bm{s}_j) + \beta_1 \cdot \text{x}_t(\bm{s}_j) + \text{w}_1(\bm{s}_j) \cdot \text{x}_t(\bm{s}_j),
\end{equation}

where $\beta_0$ is an intercept, $\eta_t$ is a temporal random effect (i.e., season-specific intercept) to account for unmodeled temporal autocorrelation in occurrence probability (only applicable if $T > 1$), $\text{w}_0(\bm{s}_j)$ is a spatial random effect to account for unmodeled spatial variation in occurrence probability, $\beta_1$ is the non-spatial effect of the covariate $\text{x}_t(\bm{s}_j)$, and $\text{w}_1(\bm{s}_j)$ is the spatially-varying effect of the covariate at each spatial location $\bm{s}_j$. 

The spatially-varying effect in the SVC model can be estimated in a variety of ways, including via generalized additive models (GAMs; \citealt{wood2006generalized}) and Gaussian processes \citep{banerjee2014hierarchical}. We focus on the latter due to their prevalence in spatial statistics, their comparatively higher predictive performance \citep{golding2016fast}, and the potential for oversmoothing of relationships with GAMs \citep{stein2014limitations}. Specifically, we have
\begin{equation}\label{w}
	\text{$\text{w}_1(\bm{s})$} \sim N(\bm{0}, \bm{C}(\bm{s}, \bm{s}', \bm{\theta})),
\end{equation}

where $\bm{C}(\bm{s}, \bm{s}', \bm{\theta})$ is a $J \times J$ covariance matrix that is a function of the distances between any pair of site coordinates $\bm{s}$ and $\bm{s}'$ and a set of parameters ($\bm{\theta}$) that govern the spatial process according to a spatial correlation function. Here we use Nearest Neighbor Gaussian Processes \citep{datta2016hierarchical} as an efficient approximation to the full Gaussian process and an exponential correlation function, which models the correlation in the spatially-varying effect of the covariate using two parameters, $\bm{\theta} = \{\sigma^2, \phi\}$, where $\sigma^2$ is a spatial variance parameter and $\phi$ is a spatial decay parameter. See \cite{doser2023SVC} for full details on SVC occupancy models.

\subsection*{How do spatially-varying coefficient models compare to simpler alternatives?}

For comparison to alternative models, consider a more generalized model of occurrence probability $\psi_t(\bm{s}_j)$ with the form
\begin{equation}\label{broadPsi}
   \text{logit}(\psi_t(\bm{s}_j)) = \beta_0 + \eta_t + \text{w}_0(\bm{s}_j) + f(\text{x}_t(\bm{s}_j), \bm{\beta}),
\end{equation}

where $\beta_0$, $\eta_t$, and $\text{w}_0(\bm{s}_j)$ are defined in (\ref{psi}), and$f(\text{x}_t(\bm{s}_j), \bm{\beta})$ is some generic function that relates the covariate $\text{x}_t(\bm{s}_j)$ to occurrence probability through a set of parameters $\bm{\beta}$. Note that $\text{w}_0(\bm{s}_j)$ and $\eta_t$ can be removed from (\ref{broadPsi}) if not applicable for a target species/data set. We consider five functional relationships to describe the species-environment relationship between $\psi_t(\bm{s}_j)$ and $\text{x}_t(\bm{s}_j)$, which vary in their ability to estimate non-linear and/or spatially-varying species environment relationships: 

\begin{enumerate}
    \item \textbf{Linear}: $f(\text{x}_t(\bm{s}_j), \bm{\beta}) = \beta_1 \cdot \text{x}_t(\bm{s}_j)$
    \item \textbf{Quadratic}: $f(\text{x}_t(\bm{s}_j), \bm{\beta}) = \beta_1 \cdot \text{x}_t(\bm{s}_j) + \beta_2 \cdot \text{x}_t^2(\bm{s}_j)$
    \item \textbf{Stratum}: $f(\text{x}_t(\bm{s}_j), \bm{\beta}) = \beta_1 \cdot \text{x}_t(\bm{s}_j) + \beta_{2, \text{STRATUM}_j} \cdot \text{x}_t(\bm{s}_j)$
    \item \textbf{Interaction}: $f(\text{x}_t(\bm{s}_j), \bm{\beta}) = \beta_{1} \cdot \text{x}_t(\bm{s}_j) + \beta_2 \cdot \text{x}_t^\ast(\bm{s}_j) \cdot \text{x}_t(\bm{s}_j)$
    \item \textbf{SVC}: $f(\text{x}_t(\bm{s}_j), \bm{\beta}) = \beta_1 \cdot \text{x}_t(\bm{s}_j) + \text{w}_1(\bm{s}_j) \cdot \text{x}_t(\bm{s}_j)$
\end{enumerate}

The linear model simply assumes a linear species-environment relationship. The quadratic model extends the linear model by allowing the species-environment relationship to peak at some optimum level and subsequently decrease as one moves farther from the optimum (i.e., a concave relationship) or peak at the extremes of the environmental variable (i.e., a convex relationship). The stratum model estimates an overall linear effect of the environmental predictor as well as  stratum-specific adjustments in the effect across a set of strata (e.g., ecoregions, management units, protected areas). Note the stratum-specific deviations could be estimated either as fixed or random effects. The interaction model similarly estimates a linear species-environment relationship, but allows for spatial variation in this relationship in the form of an interaction with a second covariate that varies across the $J$ sites ($\text{x}^\ast_t(\bm{s}_j)$). The SVC model estimates an overall linear species-environment relationship, but allows this linear relationship to vary across each site in the data set. The spatially-varying adjustment in the SVC model ($\text{w}_1(\bm{s}_j)$) serves as a local adjustment of the species-environment relationship from the overall effect $\beta_1$. The SVC model can be viewed as an extension of the stratum model, where the strata are now individual sampling sites, or as an extension of the interaction model, where the interacting ``covariate'' is now unknown and estimated as part of the model fitting process. 

\begin{figure}[ht]
    \centering
    \includegraphics[width=14cm]{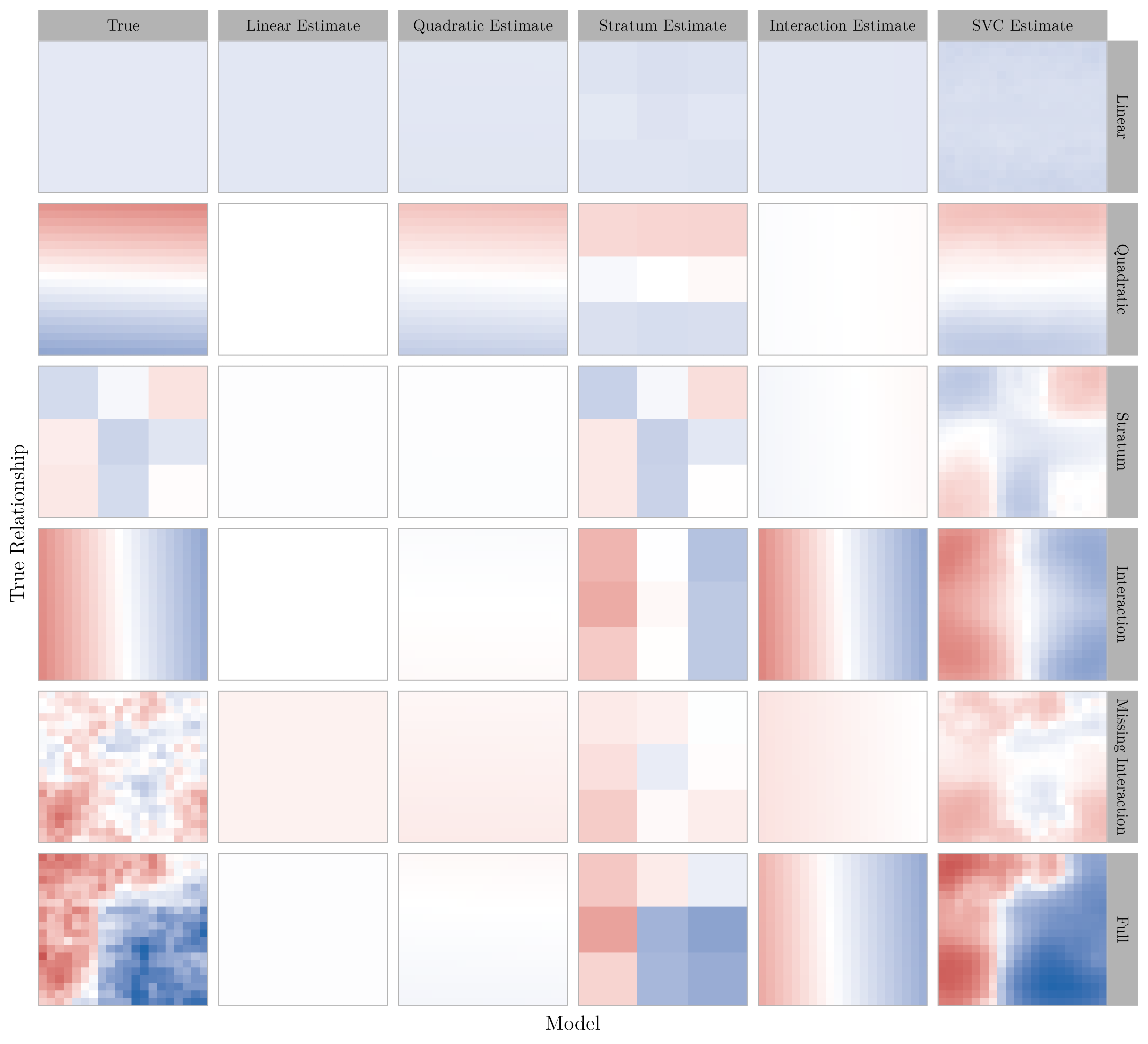}
    \caption{Estimates of a simulated species-environment relationship from different models (columns) under different patterns of the species-environment relationship (rows). Red represents a negative effect, white no effect, and blue a positive effect, with darker shades representing stronger effects. The simulated covariate varies from negative at the bottom of the grid to positive at the top of the grid. The true species-environment relationship is simulated as a linear effect (row 1), a quadratic effect (row 2), a separate linear effect across nine strata (row 3), an interaction with a second covariate that varies along the horizontal axis (row 4), an interaction with an unknown (``missing'') covariate (row 5), and the sum of all the aforementioned components (``Full'' effect, row 6). Estimates are shown from five candidate models relative to the truth (column 1), including a model with a: linear effect (column 2), quadratic effect (column 3), stratum-specific effect (column 4), interaction (column 5), and an SVC (column 6).}
    \label{fig:overviewFigure}    
\end{figure}

We performed a simulation study to compare the five aforementioned models and assess their ability to estimate species-environment relationships of different forms. Briefly, we simulated detection-nondetection data across a $20 \times 20$ grid, where occurrence probability was generated as a function of a single covariate that took negative values at the southern portion of the simulated area and positive values at the northern portion of the area. We varied the true species-environment relationship across six different functional forms (leftmost column, Figure \ref{fig:overviewFigure}): (1) linear; (2) quadratic; (3) a separate linear effect across nine strata; (4) an interaction with a second covariate that varied along the horizontal axis; (5) an interaction with an unknown (``missing'') covariate; and (6) the sum of the five aforementioned components (i.e., ``Full'' effect). We simulated 50 data sets under each functional form of the species-environment relationship, and fit the five models to each data set using the \texttt{spOccupancy} R package \citep{doser2022spoccupancy}, where the interaction model only incorporated the covariate that was assumed known, not the unknown covariate. Model performance was compared using the Widely Applicable Information Criterion (WAIC; \citealt{watanabe2010}). See Supplemental Information S1 for complete simulation details. 

The SVC model was consistently able to capture the true relationship across all six forms of the species-environment relationship, while the simpler models only performed well in a subset of scenarios (Figure \ref{fig:overviewFigure}). When the true species-environment was linear, all models yielded virtually identical estimates of the species-environment relationship that closely resembled the truth (Supplemental Information S1 Table 1). The simpler models with a pre-specified functional form (e.g., quadratic, stratum, known interaction) outperformed the SVC model according to WAIC when data were generated with the exact species-environment relationship, suggesting that when the true form of the species-environment relationship is known, simpler models should be used over the more complicated SVC model. However, when the true species-environment relationship was different from the form used to fit the model, the simpler models were limited in their ability to capture the underlying patterns (Figure \ref{fig:overviewFigure}) and performed substantially worse according to WAIC (Supplemental Information S1: Table S1). The SVC model was able to accurately capture the true pattern in the relationship even when the data were not generated with an SVC, indicating the ability of SVC models to reveal simpler functional forms of species-environment relationships when such relationships are not known prior to model fitting. Further, the SVC model drastically outperformed all other models when the species-environment relationship interacted with an unknown variable (row 5, Figure \ref{fig:overviewFigure}), or when the species-environment relationship was determined by multiple components (row 6, Figure \ref{fig:overviewFigure}).  

\section*{Case Study 1: Spatially-varying occurrence trends in eastern US forest birds}

Quantifying spatially-explicit trends can help identify areas of conservation interest (e.g., climate-change refugia) as well as provide insights on species range dynamics (e.g., trailing vs. leading edge trends). In this case study, we assessed occurrence trends of forest bird species across a ${\sim}4.04$ million km$^2$ region of the eastern US (i.e., the continental US east of the 100th meridian) from 2000-2019 ($T = 20$ years) using detection-nondetection data from the North American Breeding Bird Survey (BBS; \citealt{pardieck2020north}). We restricted our analysis to a community of 66 eastern forest bird species following the habitat classification of \cite{bateman2020north}. We subsequently assessed trends for 51 of the 66 species whose breeding ranges (derived from \citealt{birdLife}) had at least 50\% overlap with the study area (see Supplemental Information S2: Table 3 for species names). Our objectives for this case study were to (1) develop spatially-explicit maps of occurrence trends for each of the 51 species across the eastern US, and (2) compare an SVC occupancy model to three alternative models that represent simple hypotheses regarding the drivers of distributional change. Specifically, our four hypotheses and associated models were: 

\begin{enumerate}
    \item The species has a constant, linear trend across the eastern US (i.e., the linear model)
    \item The species trend varies across broad ecologically distinct strata, Bird Conservation Regions (i.e., strata model), as a result of differences in bird communities, habitat types, and management initiatives.
    \item The species trend interacts with the 30-year (1981-2010) climate normal (i.e., temperature model). If climate shapes species distributions, we would expect differences in species trends near the climatic extremes as annual temperatures become increasingly warm (e.g., positive trends at northern range boundary and negative trends at southern range boundary indicating a northward shifting distribution). 
    \item The species trend varies with a variety of interacting abiotic and biotic variables that together result in a complex spatially-varying trend across the species range (i.e., SVC model).
\end{enumerate}

We used data from $J = 1846$ BBS routes (i.e., sites) sampled at least once between 2000-2019 (mean number of sampled years per route = 15). BBS observers performed a three-minute point count survey at each of 50 stops along each route, counting all birds seen or heard within a 0.4 km radius. We summarized the data for each species at each site into $K = 5$ spatial replicates (each comprising data from 10 of the 50 stops), where each replicate took value 1 if the species was detected at any of the 10 stops in that replicate, and value 0 if the species was not detected. While such an approach has been used in previous studies with BBS data (e.g., \citealt{rushing2020migratory}), this use of spatial replicates in an occupancy modeling framework likely leads to violation of the closure assumption \citep{kendall2009cautionary}, and so we refer to our response as species-specific occurrence (or ``use'') rather than occupancy. 

For each of the 51 species, we fit a multi-season occupancy model where occurrence probability at each site $j$ in each year $t$ was modeled as 
\begin{equation}
   \text{logit}(\psi_t(\bm{s}_j)) = \beta_0 + \text{w}_0(\bm{s}_j) + \beta_1 \cdot \text{TMAX}(\bm{s}_j) + f(\text{YEAR}_t, \bm{\beta}_{\text{TREND}}) + \eta_t, 
\end{equation}

where $\beta_0$ and $\text{w}_0(\bm{s}_j)$ together represent the spatially-varying intercept, $\beta_1$ is the effect of the 30-year (1981-2010) maximum temperature climate normal on occurrence probability, $\eta_t$ is an AR(1) random year effect to accommodate residual temporal autocorrelation, and $f(\text{YEAR}_t, \bm{\beta}_{\text{TREND}})$ is the estimated trend parameter(s) that varies in form across the four models. For all models, we expressed detection probability as a function of linear and quadratic effects of year, linear and quadratic effects of survey day (to account for seasonal variation in detection probability), and linear and quadratic effects of survey replicate (to account for variability in detection probability over the time of day). For each species, we used pre-existing published ranges from BirdLife International \citep{birdLife} and only included routes that fell inside a 50 km buffer of the species range when fitting the occupancy model. Thirty-year average maximum temperature was calculated at each route from TerraClimate \citep{abatzoglou2018terraclimate}.

We compared the four models using the WAIC as an assessment of model parsimony. Using realized observations from BBS in 2021, we additionally compared each model's ability to predict (i.e., forecast) future occurrence of the species in 2021 using the area under the receiver operating characteristic curve (AUC; \citealt{hosmer2013applied}) following approaches outlined by \cite{zipkin2012evaluating}. See Supplemental Information S2 for full details. 

We fit all models using Bayesian inference with Markov chain Monte Carlo (MCMC) algorithms implemented in the \texttt{spOccupancy} R package \citep{doser2022spoccupancy, doser2023SVC}. Briefly, we specified vague prior distributions for all non-spatial parameters in the model, and weakly informative priors on the spatial parameters, with full details provided in Supplemental Information S2. We ran three chains of each model for 100,000 MCMC iterations with a burn-in period of 50,000 iterations and a thinning rate of 50, yielding 3000 posterior samples. We assessed convergence using the potential scale reduction factor (i.e., $\hat{\text{R}}$; \citealt{brooks1998}). After fitting each of the four alternative models, we predicted across the range of each species in the study area to generate maps of occurrence trends across the twenty year period. 

\subsection*{Results}

\begin{figure}
    \centering
    \includegraphics[width=15cm]{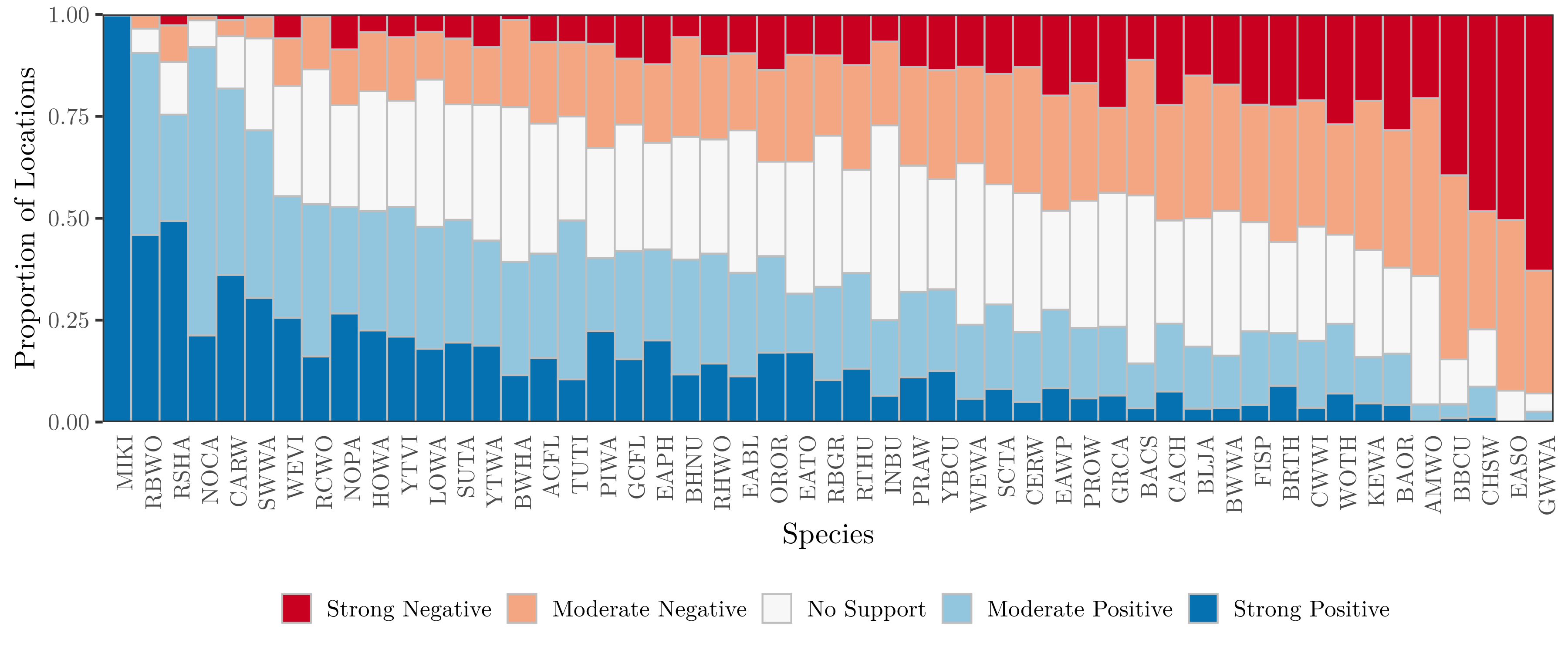}
    \caption{Summary of the spatially-varying trend of the 51 forest bird species. The height of each bar corresponds to the proportion of locations for the given species whose trend has the sign (i.e., positive, negative, no effect) and strength (i.e., strong, moderate) indicated by the color. Blue indicates support for positive trends and red indicates support for negative trends. More specifically: (1) Dark blue = Strong Positive: (P(trend $> 0$) $> 0.8$)); (2) Light blue = Moderate Positive: ($0.6 < $ P(trend $> 0$) $\leq 0.8$; (3) White = No effect: ($0.4 < $ P(trend $> 0$) $\leq 0.6$; (4) Light red = Moderate Negative: ($0.2 < $ P(trend $> 0$) $\leq 0.4$; (5) Dark red = Strong Negative: (P(trend $> 0$) $< 0.2$).}
    \label{fig:trend-summary}
\end{figure}

There was strong support for spatial variability in twenty year occurrence trends across the eastern US for the majority of the forest bird species that we included in our analysis, with many species showing both positive and negative trends across their breeding range (Figure \ref{fig:trend-summary}). The SVC model substantially outperformed (i.e., $\Delta\text{WAIC} > 2$) the linear trend model, strata model, and maximum temperature interaction model for almost all species (49, or 96\%, out of 51) according to the WAIC, indicating the additional flexibility provided by the SVC model increased model parsimony. Among the three alternative models, the BCR model performed better than the temperature model and linear model for 78\% and 90\% of species, respectively, according to the WAIC, while the temperature model generally showed less support across the 51 species, outperforming the linear model for 57\% of species. The two species with less support for the SVC model (American Woodcock and Eastern Screech Owl) had very low raw occurrence probabilities (i.e., $< 0.05$). The SVC model requires more observations than the simpler strata model to yield estimates with reasonable uncertainty, which likely contributed to the higher performance of the simpler strata model for these two species. We found more variable support for improvements in predictive performance for the SVC model relative to the three alternative models. Overall, the SVC model generally had the highest performance in predicting future species-specific occurrence (i.e., in 2021), with AUC being highest for 69\% (35 out of 51) of species (Supplemental Information S2 Table 3).

\begin{figure}
    \centering
    \includegraphics[width = 15cm]{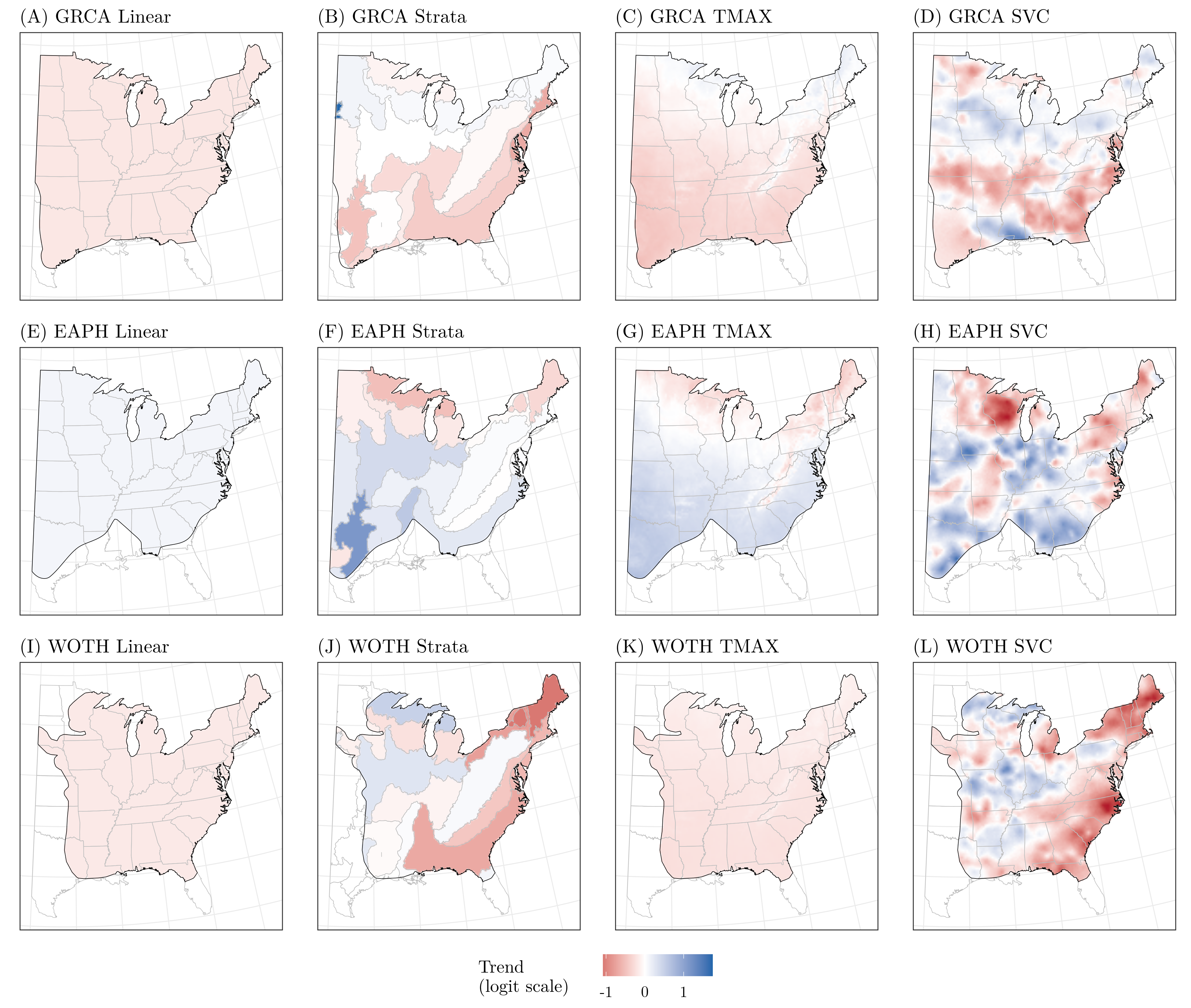}
    \caption{Median predictions of an occurrence trend from 2000-2019 for three example species from the four models: a spatial occupancy model with a constant linear trend across the species range (Linear), a spatial occupancy model with a separate trend for each Bird Conservation Region (Strata), a spatial occupancy model with a trend that interacts with 30-year average maximum temperature (TMAX), and a spatially-varying coefficient occupancy model estimating a spatially-varying trend (SVC). Panels A-D: Gray Catbird (3.79 million km$^2$); Panels E-H: Eastern Phoebe (3.55 million km$^2$); Panels I-L: Wood Thrush (3.12 million km$^2$).} 
    \label{fig:trendComparison}
\end{figure}

The SVC model revealed spatial heterogeneity in occurrence trends that was not evident in the three simpler alternative models. Figure~\ref{fig:trendComparison} shows the trend estimates for three example species (Gray Catbird (GRCA), Eastern Phoebe (EAPH), and Wood Thrush (WOTH)) from each of the four models. The SVC model revealed Gray Catbird had predominately negative trends in the southern portion of its range (except Louisiana and Mississippi) and positive or no directional trends in the northern states, indicating a potential northward shift in its range. Both the temperature model and the strata model were able to capture the general pattern of more negative trends in the southern portion of the eastern US, but they both failed to capture more fine scale variability that was revealed by the SVC model (i.e., positive trends in Louisiana and Mississippi and negative trends in northern Minnesota). We found the opposite pattern for Eastern Phoebe, where the strata and temperature models were able to adequately capture a generally positive trend in southern portions of the range and negative trend in northern portions of the range, but did not capture more fine-scale variability in trends revealed by the SVC model. Wood Thrush trends were strongly negative along the eastern portion of its range and more positive in the northwestern part of its range, which were largely captured by both the SVC and strata model. In contrast, the lack of relationship between maximum temperature and trend for Wood Thrush resulted in essentially no spatial variation in the trend from the temperature model. This clearly illustrates the lack of flexibility in the interaction models if the interacting variable does not adequately explain spatial variability in the trend.

The temperature model revealed large heterogeneity in the sign and significance of the interaction between maximum temperature and the yearly trend (Supplemental Information S2: Figure 1), providing minimal support for climatic niche position being a consistent driver of spatial variation in eastern forest bird occurrence trends. Of the 51 total species, 18 species had a significant negative interaction (i.e., trends were less positive/more negative in areas with higher temperatures), while 8 species had a significant positive interaction (i.e., trends were more positive/less negative in areas with higher temperatures). While such significant trends indicate some spatial variability in trends may be related to climatic position within a range, the large improvement in model fit (and to a lesser extent, prediction) of the SVC model compared to the temperature model suggests that this relationship is not the primary driver of spatial variability in occurrence trends for eastern forest birds. Similarly, the improved performance of the strata model compared to the constant linear trend model suggests spatial variability in trends may be partially attributed to broad-scale variation in habitat and climate conditions across Bird Conservation Regions. However, the improved performance of the SVC model compared to the strata model suggests that additional factors operating at finer spatial scales are important contributors to spatial variability in occurrence trends.

\section*{Case study 2: Effects of land cover change on the Grasshopper Sparrow}

Quantifying the effects of land cover change and resulting shifts in habitat availability on species distributions is crucial for understanding the primary drivers of large-scale avian population declines \citep{rosenberg2019decline}. Grassland birds have experienced some of the steepest population declines across all bird groups in North America, which is primarily thought to be a result of habitat loss, agricultural intensification, and increased use of pesticides in agricultural landscapes \citep{stanton2018analysis}. However, the effects of shifts in land cover on grassland bird species is unlikely to be constant across space as a result of complex interactive effects with local climate, farmland management practices (e.g., pesticide use), or predation pressure and other stressors. In this case study, we demonstrate the ability of SVC models to provide insight on the spatially-varying effects of habitat change on the distribution of Grasshopper Sparrow (\textit{Ammodramus savannarum}) across the continental US from 1970-2019. We again use data from the North American BBS \citep{pardieck2020north}, collected over the 50-year period at $J = 2542$ routes within the range of the Grasshopper Sparrow (derived from BirdLife International \citep{birdLife} as described in Case Study 1). Our objectives for this case study were to (1) develop spatially-explicit maps of the effect of change in grassland area and cropland area on Grasshopper Sparrow occurrence; and (2) compare a series of SVC and simpler alternative models to generate and test hypotheses regarding the drivers of spatial variability in the effects of habitat change.

We summarized the BBS data into replicated detection-nondetection data in the same manner as in Case Study 1, using $K = 5$ spatial replicates at each BBS route to model route-level occurrence across the sparrow's range. We calculated annual amount of grassland area (natural grassland cover class) and cropland area (combined cropland and hay/pasture cover classes) within 1km of each BBS route using data from the USGS EROS (Earth Resources Observation and Science) Center \citep{sohl2016modeled}. We calculated annual deviations in grassland and cropland area by subtracting the 50-year average value at each site from each yearly value to explicitly assess effects of temporal change in habitat separately from spatial variation in habitat availability \citep{clement2019partitioning, saunders2022unraveling}. 

We hypothesized that Grasshopper Sparrow occurrence would be positively associated with grassland area change and that the effect would vary spatially as a result of: (1) a positive interaction with the average amount of grassland area given Grasshopper Sparrow's preference for landscapes comprised of large amounts of contiguous grassland \citep{shaffer2021}; (2) an interaction with temperature such that effects of grassland area change are strongest near the climatic extremes of the species range (i.e., range boundaries); and (3) additional interactions with fine-scale habitat quality (e.g., grassland height, amount of litter; \citealt{shaffer2021}) and management actions that were not available as covariates for the model. We expected Grasshopper Sparrow occurrence to be negatively related to cropland cover change across the Great Plains region, as an increase in cropland cover in this region would likely correspond to a decrease in grassland area (the dominant land cover type in this region; Supplemental Information S3: Figure 1). However, in areas with minimal grassland (i.e., east of the 100th meridian), we predicted a positive association between Grasshopper Sparrow occurrence and cropland area, as Grasshopper Sparrow occurrence has previously been associated with hay-fields and cultivated fields when native grassland area is limited \citep{shaffer2021}.

We fit five candidate models that varied in the functional forms of the effects of grassland area change and cropland area change to test our hypotheses (full details in Supplemental Information S3). Specifically, our five models consisted of: (1) a linear model with constant, linear effects of grassland and cropland change; (2) a habitat interaction model with linear effects of grassland and cropland change, an interaction of grassland change with 50-year average grassland area, and an interaction of cropland change with 50-year average cropland area; (3) a temperature interaction model with linear effects of grassland and cropland change that both also had an interaction with average temperature conditions (i.e., 30-year maximum temperature calculated from TerraClimate as in Case Study 1); (4) an SVC model; (5) the ``full'' model that contained SVCs, interactive effects of temperature, and interactive effects of 50-year average land-cover amount. 

For each of the five candidate models, we fit a Bayesian multi-season occupancy model using the \texttt{spOccupancy} R package \citep{doser2022spoccupancy}. Detection probability was modeled consistently across the five models as a function of linear and quadratic ordinal date, a random effect of year, and a separate intercept of survey replicate to account for variability in detection probability across the five spatial replicates within a BBS route. Given our focus on inference of the species-environment relationships, we compared candidate models using the WAIC. Prior distributions were either vague or weakly informative (Supplemental Information S3). For each model, we ran three chains for 100,000 MCMC iterations with a burn-in period of 50,000 iterations and a thinning rate of 50, yielding 3000 posterior samples. Convergence was assessed using the potential scale reduction factor and visual assessment of traceplots using the \texttt{coda} package \citep{coda}. 

\subsection*{Results}

\begin{figure}
    \centering
    \includegraphics[width = 15cm]{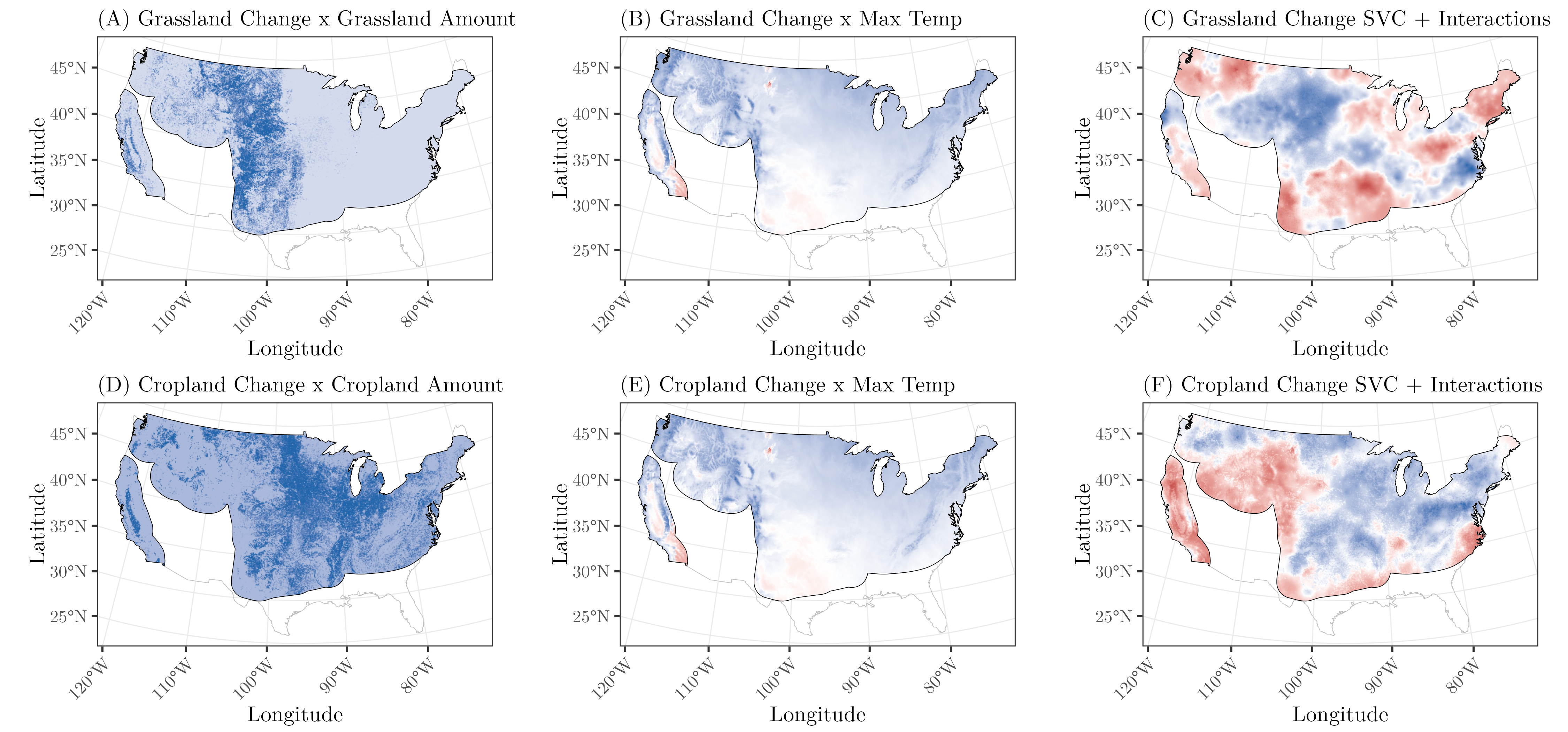}
    \caption{Median predictions of the effects of grassland change (top row) and cropland change (bottom row) on occurrence of Grasshopper Sparrow (\textit{Ammodramus savannarum}) from three of the five candidate models. Panels (A) and (D) show estimates from a model with an interaction between land-cover change and average land-cover area over the 50 year period. Panels (B) and (E) show estimates from a model with an interaction between land-cover change and 30-year average maximum temperature. Panels (C) and (F) show estimates from a model with spatially-varying coefficients for land-cover change and interactions with average land-cover area and maximum temperature. Blue indicates a positive effect, white indicates no effect, and red indicates a negative effect.} 
    \label{fig:caseStudy2}
\end{figure}

We found strong support for spatial variability in the effects of grassland and cropland cover change, with all models that included an SVC and/or an interaction substantially outperforming (i.e., $\Delta$WAIC $>2$) the model with constant effects (Supplemental Information S3: Table 1). The temperature interaction model outperformed the habitat interaction model ($\Delta\text{WAIC} = 18.65$), indicating maximum temperature was more important in explaining spatial variability in the effect of habitat change than the amount of habitat. Noticeably, including an SVC for the effect of grassland and cropland change reduced WAIC (i.e., $\Delta\text{WAIC} = 774.90$) substantially more than either of the interaction models compared to the constant model ($\Delta\text{WAIC} = 20.69$ for the habitat interaction model and $\Delta\text{WAIC} = 39.34$ for the temperature interaction model). The model including SVCs, a habitat interaction, and a temperature interaction slightly outperformed the model with only SVCs ($\Delta\text{WAIC} = 3.32$). Altogether, these results suggest that interactions with temperature and habitat explain some spatial variability in the effect of habitat change on grassland bird occurrence, but most of the variation in these effects is the result of unexplained spatial variation that is accounted for by the SVCs. Maps of the predicted effects of grassland and cropland change from the different candidate models reveal that models with the SVC capture far more spatial variation in the effects of habitat change than do models without SVCs (Figure \ref{fig:caseStudy2}).

The effects estimated under the candidate models revealed mixed support for our hypotheses. The best performing model revealed a range of positive and negative effects of habitat change across the breeding range of the Grasshopper Sparrow. As predicted, the effect of grassland change was strongly positive in the Northern Great Plains (where grassland area is high), suggesting that in heavily grassland-dominated landscapes, decreasing amounts of grassland would result in declines in Grasshopper Sparrow occurrence probability. This is further supported by the habitat interaction model, which revealed a positive interaction between grassland change and average grassland area (Figure \ref{fig:caseStudy2}A). Surprisingly, we found near zero or negative effects of grassland area change in the Southern Great Plains, indicating increasing grassland in this area would result in no effect or even declines of Grasshopper Sparrow occurrence probability. Given the relatively high amount of grassland and rangeland area in this region (Supplemental Information S3 Figure 1), this could indicate a regional peak in the optimal amount of grassland area for Grasshopper Sparrow occurrence probability, which is in line with previous work showing that the preferred grassland size of Grasshopper Sparrows varies across ecoregions \citep{johnson2001area}. The temperature interaction model revealed this pattern was partially related to a negative interaction with maximum temperature (Figure \ref{fig:caseStudy2}B; \citealt{gorzo2016using}). Grasshopper Sparrow occurrence probability was negatively related to cropland area along most of its southern range boundary, which was partially attributable to a negative interaction with maximum temperature (Figure \ref{fig:caseStudy2}). Alternatively, the effect of cropland change was generally positive throughout the Northeast and Midwest (Figure \ref{fig:caseStudy2}), which was partially related to a positive interaction between cropland change and cropland area (Figure \ref{fig:caseStudy2}D). In the northeastern and midwestern states, where few native grasslands remain, the positive effect of cropland area change indicates that increases in cropland area would result in increases in Grasshopper Sparrow occurrence probability. Further, the northeastern US is largely dominated by forest, and thus increases in cropland cover are likely associated with declines in forest cover, which may partly explain the positive effect of cropland cover change, as Grasshopper Sparrow completely avoids forest \citep{grant2004tree}. While such assessments are speculative, these insights are only possible because of the power of SVC models to reveal fine-scale, multifaceted species-environment relationships, which in turn can be used to inform local and regional management and conservation priorities.

\section*{Discussion}

Accounting for complex spatially-varying and/or nonlinear species-environment relationships is increasingly important as the scope of ecological research expands in spatial and temporal extent \citep{rollinson2021working}. Widely used statistical methods, such as interactions, stratification, and nonlinear models can partially account for such patterns, but they are limited in their flexibility to estimate spatially-varying species-environment relationships that arise from multiple interacting factors that themselves vary spatially, especially if such factors are not available as covariates. Here we discussed the use of spatially-varying coefficients (SVCs) in SDMs, which provide a powerful approach for modeling nonlinear and/or spatially-varying species environment relationships within a hierarchical GLM framework that can simultaneously address observational biases common in both wildlife and plant (e.g., \citealt{chen2013imperfect}) datasets. Using simulations and two case studies on birds in the US, we highlighted the inferential benefits of SVC SDMs to generate and test ecological hypotheses regarding the factors driving spatial variability in estimated relationships and/or occurrence trends.

Our simulation study revealed that SVC SDMs can accurately capture complex, spatially-varying species-environment relationships under different forms, while they can also reveal more simple species-environment relationships (e.g., linear, quadratic; Figure \ref{fig:overviewFigure}) if such additional complexity is not supported by the data. Thus, when little is known regarding the form of the species-environment relationship prior to model fitting, SVC SDMs can be used to generate hypotheses on the true form of the relationship and what abiotic and/or ecological factors influence the relationship. When the true species-environment relationship is known \textit{a priori}, simpler parametric GLMs will likely outperform SVC SDMs according to information criteria based on the principle of parsimony, given the increased complexity of SVC models. In such situations, our simulation study suggests that SVC SDMs will not generate erroneous conclusions, but rather will reveal the simpler, parametric form of the true relationship. When working across macroscales, it is unlikely that all interacting variables are known and/or available prior to model fitting, in which case SVC SDMs will outperform simpler alternatives (rows 5, 6; Figure \ref{fig:overviewFigure}) and help reveal the ecological drivers of such patterns. 

As shown in the two empirical case studies, a key benefit of SVC SDMs is the ability to test and subsequently generate hypotheses regarding the drivers of spatial variability in species-environment relationships and occurrence trends. When assessing spatial variability in species-environment relationships and/or trends, we recommend comparing SVC SDMs with simpler parametric SDMs that represent explicit hypotheses, as such comparisons can reveal the amount of support for different drivers of spatially-varying effects/trends \citep{pease2022drives, sultaire2022spatial}. For example, in the eastern forest bird case study, the temperature model revealed a significant negative interaction of trend and breeding season maximum temperature for 18 species and a significant positive interaction for eight species (Supplemental Information S3 Figure 1). However, the SVC model was the best-performing model for all 26 species with significant temperature interactions, suggesting that while breeding season temperature often explains some variation in occurrence trends, there are additional factors (e.g., change in habitat quality, climate change) that are important in explaining fine-scale variability in occurrence trends. 

Whether SVC models improve predictive performance over models that only include a spatially-varying intercept is an ongoing statistical question. In our eastern forest bird case study, the SVC model provided relatively minor improvements in predictive performance compared to the simpler models when forecasting occurrence probability in 2021. All four models included a spatial random effect to account for spatial variability in occurrence probability, and given the likely small changes in the forest bird distributions from 2019 (the last year in the modeled data set) to 2021, they all had similar abilities to predict occurrence probabilities across the study region. SVC models in other ecological and natural resource applications have shown mixed results regarding their predictive benefits compared to models with only a spatially-varying intercept: some studies found improved predictive performance of SVC models \citep{sultaire2022spatial, may2023spatially}, while others showed improvements that vary depending on the species \citep{pease2022exploring, doser2023SVC} or region \citep{babcock2015lidar}. Regardless, we echo the statements of \cite{thorson2023spatially} that the primary benefits of SVC SDMs relate to their improved ability to test and generate hypotheses as well as answer relevant ecological questions regarding spatial variability in species-environment relationships and trends.

In addition to theoretical contributions, the results from SVC SDMs (e.g., estimates of species-environment relationships/trends across a species range) could be applied to multi-scale conservation and management decisions. For example, in the Grasshopper Sparrow case study we found that loss of grassland area is most likely to have the largest impact (i.e. resulting in declines in occurrence probabilities) in the Northern Great Plains, emphasizing the importance of providing large, contiguous patches of natural grassland to prevent further declines of this species in the region \citep{shaffer2021}. By performing similar analyses for multiple grassland bird species, SVC SDM outputs could be used together as inputs for spatial prioritization analyses. This could offer major improvements in reserve design and help resource managers identify the exact locations where habitat restoration may be most beneficial to the overall bird community \citep{grand2019future}. Alternatively, estimates of species trends serve as the foundation for assigning conservation status to species of greatest conservation need. Spatially-varying trends from SVC SDMs, like those generated in the eastern forest bird case study, could be used in such assessments across local (i.e., state-level), regional (i.e., ecoregion), and continental scales \citep{smith2023spatially}. Ultimately, this could improve our understanding of how and why conservation actions in different regions lead to variable outcomes. Providing managers with multi-scale occurrence trends allows for tailored action plans, and thus, more effective recovery strategies for species of conservation concern. Estimation of local trends with SVC SDMs can also improve predictions of species distribution changes \citep{barnett2021improving} in response to invasive species \citep{thorson2023spatially} and future climate and/or land-use changes \citep{gonthier2014biodiversity}.

While other statistical (e.g., GAMs) and machine learning (e.g., random forests, MaxEnt) approaches are commonly used in ecology to account for complex species-environment relationships, the Bayesian spatially-varying coefficient models described here are an attractive alternative as they (1) do not require \textit{a priori} knowledge of interacting variables; (2) can readily provide uncertainty measures associated with all estimates; and (3) are easily embedded in hierarchical modeling frameworks (i.e., occupancy models) used to address observational biases prevalent in ecological data. Nevertheless, the flexibility provided by Bayesian SVC SDMs can lead to computational and practical difficulties in their implementations. While the Bayesian framework provides full uncertainty propagation into all estimates and predictions, models can take substantial time to run. For example, the full SVC model for the Grasshopper Sparrow case study with a data set comprised of nearly 400,000 observations took approximately 10 hours to run a single MCMC chain of 100,000 samples using \texttt{spOccupancy} \citep{doser2022spoccupancy}. The \texttt{R} packages \texttt{VAST} \citep{thorson2019guidance} and \texttt{sdmTMB} \citep{anderson2022sdmtmb} provide maximum likelihood (i.e., frequentist) alternatives to fit SVC SDMs. They are substantially faster, but these models do not explicitly account for imperfect detection. Additionally, the ability of SVC SDMs to estimate complex spatially-varying species-environment relationships can require large sample sizes to achieve reasonable levels of uncertainty compared to simpler alternatives (i.e., stratification, interactions). This is particularly true when working with detection-nondetection data, which provide relatively little information to estimate SVCs compared to count (e.g., abundance) or continuous (e.g., biomass) data sources used in many SDMs. In Supplemental Information S4, we provide additional simulation studies that give insights on how the reliability of SVC estimates scales with the number of spatial locations and number of seasons in the data set. When sample sizes are limited, simpler approaches like stratification or interactions may be more useful options to yield estimates of species-environment relationships without considerable uncertainty. Lastly, confounding can occur between the estimated spatially-varying intercept and spatially-varying coefficients, especially when working with modestly-sized data sets (e.g., 500 data points), which could potentially lead to misleading conclusions. However, when estimating SVCs for covariates that vary across time (e.g., a temporal trend or habitat change as in our two case studies), confounding is minimized due to to the added temporal component of multi-season data. We have found that recent guidelines for minimizing spatial confounding and understanding its effects in spatially-explicit SDMs is applicable to SVC SDMs \citep{makinen2022spatial}, although further research is needed to understand when such confounding may occur and how to best mitigate it.  

Spatial variability in species-environment relationships is prevalent throughout ecology \citep{rollinson2021working} as a result of complex interactions with abiotic and biotic variables, which are rarely all known or available to be measured prior to statistical analysis. As we demonstrate in this study, the use of spatially-varying coefficients in species distribution models can help elucidate the environmental factors that drive species distributional dynamics across both local and broad spatial scales. This provides an improved ability to test ecological hypotheses and inform multi-scale conservation and management initiatives. When fitting SDMs across macroscales, we encourage the comparison of SVC SDMs with simpler alternatives as a means of advancing our understanding of the drivers of species-environment relationships across space.

\bibliographystyle{apalike}
\bibliography{references}

\begin{thebibliography}{}

\bibitem[Abatzoglou et~al., 2018]{abatzoglou2018terraclimate}
Abatzoglou, J.~T., Dobrowski, S.~Z., Parks, S.~A., and Hegewisch, K.~C. (2018).
\newblock {TerraClimate, a high-resolution global dataset of monthly climate
  and climatic water balance from 1958--2015}.
\newblock {\em Scientific Data}, 5(1):1--12.

\bibitem[Amburgey et~al., 2018]{amburgey2018range}
Amburgey, S.~M., Miller, D.~A., Campbell~Grant, E.~H., Rittenhouse, T.~A.,
  Benard, M.~F., Richardson, J.~L., Urban, M.~C., Hughson, W., Brand, A.~B.,
  Davis, C.~J., et~al. (2018).
\newblock Range position and climate sensitivity: The structure of
  among-population demographic responses to climatic variation.
\newblock {\em Global Change Biology}, 24(1):439--454.

\bibitem[Anderson et~al., 2022]{anderson2022sdmtmb}
Anderson, S.~C., Ward, E.~J., English, P.~A., and Barnett, L.~A. (2022).
\newblock {sdmTMB: an R package for fast, flexible, and user-friendly
  generalized linear mixed effects models with spatial and spatiotemporal
  random fields}.
\newblock {\em bioRxiv}.

\bibitem[Babcock et~al., 2015]{babcock2015lidar}
Babcock, C., Finley, A.~O., Bradford, J.~B., Kolka, R., Birdsey, R., and Ryan,
  M.~G. (2015).
\newblock Lidar based prediction of forest biomass using hierarchical models
  with spatially varying coefficients.
\newblock {\em Remote Sensing of Environment}, 169:113--127.

\bibitem[Babcock et~al., 2016]{babcock2016modeling}
Babcock, C., Finley, A.~O., Cook, B.~D., Weiskittel, A., and Woodall, C.~W.
  (2016).
\newblock Modeling forest biomass and growth: Coupling long-term inventory and
  lidar data.
\newblock {\em Remote Sensing of Environment}, 182:1--12.

\bibitem[Banerjee et~al., 2014]{banerjee2014hierarchical}
Banerjee, S., Carlin, B.~P., and Gelfand, A.~E. (2014).
\newblock {\em Hierarchical modeling and analysis for spatial data}.
\newblock Chapman and Hall/CRC.

\bibitem[Barnett et~al., 2021]{barnett2021improving}
Barnett, L.~A., Ward, E.~J., and Anderson, S.~C. (2021).
\newblock Improving estimates of species distribution change by incorporating
  local trends.
\newblock {\em Ecography}, 44(3):427--439.

\bibitem[Bateman et~al., 2020]{bateman2020north}
Bateman, B.~L., Wilsey, C., Taylor, L., Wu, J., LeBaron, G.~S., and Langham, G.
  (2020).
\newblock {North American birds require mitigation and adaptation to reduce
  vulnerability to climate change}.
\newblock {\em Conservation Science and Practice}, 2(8):e242.

\bibitem[{BirdLife International}, 2021]{birdLife}
{BirdLife International} (2021).
\newblock {Handbook of the Birds of the World 2021. Bird species distribution
  maps of the world, Ver. 2021.}
\newblock \url{http://datazone.birdlife.org/species/requestdis}.

\bibitem[Bled et~al., 2013]{bled2013modeling}
Bled, F., Sauer, J., Pardieck, K., Doherty, P., and Royle, J.~A. (2013).
\newblock {Modeling trends from North American Breeding Bird Survey data: A
  spatially explicit approach}.
\newblock {\em PLoS One}, 8(12):e81867.

\bibitem[Brooks and Gelman, 1998]{brooks1998}
Brooks, S.~P. and Gelman, A. (1998).
\newblock General methods for monitoring convergence of iterative simulations.
\newblock {\em Journal of Computational and Graphical Statistics},
  7(4):434--455.

\bibitem[Chen et~al., 2013]{chen2013imperfect}
Chen, G., K{\'e}ry, M., Plattner, M., Ma, K., and Gardner, B. (2013).
\newblock Imperfect detection is the rule rather than the exception in plant
  distribution studies.
\newblock {\em Journal of Ecology}, 101(1):183--191.

\bibitem[Clark and Luis, 2020]{clark2020nonlinear}
Clark, T. and Luis, A.~D. (2020).
\newblock Nonlinear population dynamics are ubiquitous in animals.
\newblock {\em Nature Ecology \& Evolution}, 4(1):75--81.

\bibitem[Clement et~al., 2019]{clement2019partitioning}
Clement, M.~J., Nichols, J.~D., Collazo, J.~A., Terando, A.~J., Hines, J.~E.,
  and Williams, S.~G. (2019).
\newblock Partitioning global change: Assessing the relative importance of
  changes in climate and land cover for changes in avian distribution.
\newblock {\em Ecology and Evolution}, 9(4):1985--2003.

\bibitem[Crossley et~al., 2021]{crossley2021complex}
Crossley, M.~S., Smith, O.~M., Davis, T.~S., Eigenbrode, S.~D., Hartman, G.~L.,
  Lagos-Kutz, D., Halbert, S.~E., Voegtlin, D.~J., Moran, M.~D., and Snyder,
  W.~E. (2021).
\newblock Complex life histories predispose aphids to recent abundance
  declines.
\newblock {\em Global Change Biology}, 27(18):4283--4293.

\bibitem[Datta et~al., 2016]{datta2016hierarchical}
Datta, A., Banerjee, S., Finley, A.~O., and Gelfand, A.~E. (2016).
\newblock {Hierarchical nearest-neighbor Gaussian process models for large
  geostatistical datasets}.
\newblock {\em Journal of the American Statistical Association},
  111(514):800--812.

\bibitem[Doser et~al., 2022]{doser2022spoccupancy}
Doser, J.~W., Finley, A.~O., K{\'e}ry, M., and Zipkin, E.~F. (2022).
\newblock {spOccupancy: An R package for single-species, multi-species, and
  integrated spatial occupancy models}.
\newblock {\em Methods in Ecology and Evolution}, 13(8):1670--1678.

\bibitem[Doser et~al., 2023]{doser2023SVC}
Doser, J.~W., Finley, A.~O., Saunders, S.~P., K\'ery, M., Weed, A.~S., and
  Zipkin, E.~F. (2023).
\newblock Modeling complex species-environment relationships through
  spatially-varying coefficient occupancy models.
\newblock {\em arXiv preprint}.

\bibitem[Ethier et~al., 2017]{ethier2017spatiotemporal}
Ethier, D.~M., Koper, N., and Nudds, T.~D. (2017).
\newblock Spatiotemporal variation in mechanisms driving regional-scale
  population dynamics of a threatened grassland bird.
\newblock {\em Ecology and Evolution}, 7(12):4152--4162.

\bibitem[Golding and Purse, 2016]{golding2016fast}
Golding, N. and Purse, B.~V. (2016).
\newblock {Fast and flexible Bayesian species distribution modelling using
  Gaussian processes}.
\newblock {\em Methods in Ecology and Evolution}, 7(5):598--608.

\bibitem[Gonthier et~al., 2014]{gonthier2014biodiversity}
Gonthier, D.~J., Ennis, K.~K., Farinas, S., Hsieh, H.-Y., Iverson, A.~L.,
  Bat{\'a}ry, P., Rudolphi, J., Tscharntke, T., Cardinale, B.~J., and Perfecto,
  I. (2014).
\newblock Biodiversity conservation in agriculture requires a multi-scale
  approach.
\newblock {\em Proceedings of the Royal Society B: Biological Sciences},
  281(1791):20141358.

\bibitem[Gorzo et~al., 2016]{gorzo2016using}
Gorzo, J.~M., Pidgeon, A.~M., Thogmartin, W.~E., Allstadt, A.~J., Radeloff,
  V.~C., Heglund, P.~J., and Vavrus, S.~J. (2016).
\newblock {Using the North American Breeding Bird Survey to assess broad-scale
  response of the continent's most imperiled avian community, grassland birds,
  to weather variability}.
\newblock {\em The Condor: Ornithological Applications}, 118(3):502--512.

\bibitem[Grand et~al., 2019]{grand2019future}
Grand, J., Wilsey, C., Wu, J.~X., and Michel, N.~L. (2019).
\newblock {The future of North American grassland birds: Incorporating
  persistent and emergent threats into full annual cycle conservation
  priorities}.
\newblock {\em Conservation Science and Practice}, 1(4):e20.

\bibitem[Grant et~al., 2004]{grant2004tree}
Grant, T.~A., Madden, E., and Berkey, G.~B. (2004).
\newblock Tree and shrub invasion in northern mixed-grass prairie: implications
  for breeding grassland birds.
\newblock {\em Wildlife Society Bulletin}, 32(3):807--818.

\bibitem[Guisan et~al., 2002]{guisan2002generalized}
Guisan, A., Edwards~Jr, T.~C., and Hastie, T. (2002).
\newblock Generalized linear and generalized additive models in studies of
  species distributions: setting the scene.
\newblock {\em Ecological Modelling}, 157(2-3):89--100.

\bibitem[Guisan et~al., 2017]{guisan2017habitat}
Guisan, A., Thuiller, W., and Zimmermann, N.~E. (2017).
\newblock {\em Habitat suitability and distribution models: with applications
  in R}.
\newblock Cambridge University Press.

\bibitem[Guisan and Zimmermann, 2000]{guisan2000predictive}
Guisan, A. and Zimmermann, N.~E. (2000).
\newblock Predictive habitat distribution models in ecology.
\newblock {\em Ecological Modelling}, 135(2-3):147--186.

\bibitem[Hosmer~Jr et~al., 2013]{hosmer2013applied}
Hosmer~Jr, D.~W., Lemeshow, S., and Sturdivant, R.~X. (2013).
\newblock {\em Applied logistic regression}, volume 398.
\newblock John Wiley \& Sons.

\bibitem[Hostetler and Chandler, 2015]{hostetler2015improved}
Hostetler, J.~A. and Chandler, R.~B. (2015).
\newblock Improved state-space models for inference about spatial and temporal
  variation in abundance from count data.
\newblock {\em Ecology}, 96(6):1713--1723.

\bibitem[Johnson and Igl, 2001]{johnson2001area}
Johnson, D.~H. and Igl, L.~D. (2001).
\newblock Area requirements of grassland birds: a regional perspective.
\newblock {\em The Auk}, 118(1):24--34.

\bibitem[Kellner and Swihart, 2014]{kellner2014accounting}
Kellner, K.~F. and Swihart, R.~K. (2014).
\newblock Accounting for imperfect detection in ecology: a quantitative review.
\newblock {\em PloS One}, 9(10):e111436.

\bibitem[Kendall and White, 2009]{kendall2009cautionary}
Kendall, W.~L. and White, G.~C. (2009).
\newblock A cautionary note on substituting spatial subunits for repeated
  temporal sampling in studies of site occupancy.
\newblock {\em Journal of Applied Ecology}, 46(6):1182--1188.

\bibitem[K{\'e}ry, 2011]{kery2011towards}
K{\'e}ry, M. (2011).
\newblock Towards the modelling of true species distributions.
\newblock {\em Journal of Biogeography}, 38(4):617--618.

\bibitem[Liaw et~al., 2002]{liaw2002classification}
Liaw, A., Wiener, M., et~al. (2002).
\newblock {Classification and regression by randomForest}.
\newblock {\em {R news}}, 2(3):18--22.

\bibitem[MacKenzie et~al., 2002]{mackenzie2002}
MacKenzie, D.~I., Nichols, J.~D., Lachman, G.~B., Droege, S., Royle, J.~A., and
  Langtimm, C.~A. (2002).
\newblock Estimating site occupancy rates when detection probabilities are less
  than one.
\newblock {\em Ecology}, 83(8):2248--2255.

\bibitem[MacKenzie et~al., 2017]{mackenzie2017occupancy}
MacKenzie, D.~I., Nichols, J.~D., Royle, J.~A., Pollock, K.~H., Bailey, L.~L.,
  and Hines, J.~E. (2017).
\newblock {\em Occupancy estimation and modeling: inferring patterns and
  dynamics of species occurrence}.
\newblock Elsevier.

\bibitem[M{\"a}kinen et~al., 2022]{makinen2022spatial}
M{\"a}kinen, J., Numminen, E., Niittynen, P., Luoto, M., and Vanhatalo, J.
  (2022).
\newblock Spatial confounding in bayesian species distribution modeling.
\newblock {\em Ecography}, 2022(11):e06183.

\bibitem[May et~al., 2023]{may2023spatially}
May, P., McConville, K.~S., Moisen, G.~G., Bruening, J., and Dubayah, R.
  (2023).
\newblock A spatially varying model for small area estimates of biomass density
  across the contiguous united states.
\newblock {\em Remote Sensing of Environment}, 286:113420.

\bibitem[Meehan et~al., 2019]{meehan2019spatial}
Meehan, T.~D., Michel, N.~L., and Rue, H. (2019).
\newblock {Spatial modeling of Audubon Christmas Bird Counts reveals fine-scale
  patterns and drivers of relative abundance trends}.
\newblock {\em Ecosphere}, 10(4):e02707.

\bibitem[Miller, 2012]{miller2012species}
Miller, J.~A. (2012).
\newblock Species distribution models: Spatial autocorrelation and
  non-stationarity.
\newblock {\em Progress in Physical Geography}, 36(5):681--692.

\bibitem[Osborne and Su{\'a}rez-Seoane, 2002]{osborne2002should}
Osborne, P.~E. and Su{\'a}rez-Seoane, S. (2002).
\newblock Should data be partitioned spatially before building large-scale
  distribution models?
\newblock {\em Ecological Modelling}, 157(2-3):249--259.

\bibitem[Pardieck et~al., 2020]{pardieck2020north}
Pardieck, K., Ziolkowski~Jr, D., Lutmerding, M., Aponte, V., and Hudson, M.-A.
  (2020).
\newblock {North American breeding bird survey dataset 1966--2019}.
\newblock {\em U.S. Geological Survey data release,
  https://doi.org/10.5066/P9J6QUF6}.

\bibitem[Pease et~al., 2022a]{pease2022exploring}
Pease, B.~S., Pacifici, K., and Kays, R. (2022a).
\newblock Exploring spatial nonstationarity for four mammal species reveals
  regional variation in environmental relationships.
\newblock {\em Ecosphere}, 13(8):e4166.

\bibitem[Pease et~al., 2022b]{pease2022drives}
Pease, B.~S., Pacifici, K., Kays, R., and Reich, B. (2022b).
\newblock What drives spatially varying ecological relationships in a
  wide-ranging species?
\newblock {\em Diversity and Distributions}.

\bibitem[Phillips et~al., 2006]{phillips2006maximum}
Phillips, S.~J., Anderson, R.~P., and Schapire, R.~E. (2006).
\newblock Maximum entropy modeling of species geographic distributions.
\newblock {\em Ecological Modelling}, 190(3-4):231--259.

\bibitem[Plummer et~al., 2006]{coda}
Plummer, M., Best, N., Cowles, K., and Vines, K. (2006).
\newblock {CODA: Convergence Diagnosis and Output Analysis for MCMC}.
\newblock {\em R News}, 6(1):7--11.

\bibitem[Reich et~al., 2018]{reich2018integrating}
Reich, B.~J., Pacifici, K., and Stallings, J.~W. (2018).
\newblock Integrating auxiliary data in optimal spatial design for species
  distribution modelling.
\newblock {\em Methods in Ecology and Evolution}, 9(6):1626--1637.

\bibitem[Rollinson et~al., 2021]{rollinson2021working}
Rollinson, C.~R., Finley, A.~O., Alexander, M.~R., Banerjee, S., Dixon~Hamil,
  K.-A., Koenig, L.~E., Locke, D.~H., DeMarche, M.~L., Tingley, M.~W., Wheeler,
  K., et~al. (2021).
\newblock Working across space and time: nonstationarity in ecological research
  and application.
\newblock {\em Frontiers in Ecology and the Environment}, 19(1):66--72.

\bibitem[Rosenberg et~al., 2019]{rosenberg2019decline}
Rosenberg, K.~V., Dokter, A.~M., Blancher, P.~J., Sauer, J.~R., Smith, A.~C.,
  Smith, P.~A., Stanton, J.~C., Panjabi, A., Helft, L., Parr, M., et~al.
  (2019).
\newblock {Decline of the North American avifauna}.
\newblock {\em Science}, 366(6461):120--124.

\bibitem[Rushing et~al., 2020]{rushing2020migratory}
Rushing, C.~S., Royle, J.~A., Ziolkowski~Jr, D.~J., and Pardieck, K.~L. (2020).
\newblock Migratory behavior and winter geography drive differential range
  shifts of eastern birds in response to recent climate change.
\newblock {\em Proceedings of the National Academy of Sciences},
  117(23):12897--12903.

\bibitem[Saunders et~al., 2022]{saunders2022unraveling}
Saunders, S.~P., Meehan, T.~D., Michel, N.~L., Bateman, B.~L., DeLuca, W.,
  Deppe, J.~L., Grand, J., LeBaron, G.~S., Taylor, L., Westerkam, H., et~al.
  (2022).
\newblock Unraveling a century of global change impacts on winter bird
  distributions in the eastern united states.
\newblock {\em Global Change Biology}, 28:2221--2235.

\bibitem[Shaffer et~al., 2021]{shaffer2021}
Shaffer, J.~A., Igl, L.~D., Johnson, D.~H., Sondreal, M.~L., Goldade, C.~M.,
  Nenneman, M.~P., Wooten, T.~L., and Euliss, B.~R. (2021).
\newblock {The effects of management practices on grassland birds—Grasshopper
  Sparrow (Ammodramus savannarum)}.
\newblock {\em U.S. Geological Survey Professional Paper}.

\bibitem[Smith et~al., 2023]{smith2023spatially}
Smith, A.~C., Binley, A., Daly, L., Edwards, B.~P., Ethier, D., Frei, B., Iles,
  D., Meehan, T.~D., Michel, N.~L., and Smith, P.~A. (2023).
\newblock {Spatially explicit Bayesian hierarchical models for avian population
  status and trends}.
\newblock {\em EcoEvoRxiv}.

\bibitem[Smith and Edwards, 2021]{smith2021north}
Smith, A.~C. and Edwards, B.~P. (2021).
\newblock {North American Breeding Bird Survey status and trend estimates to
  inform a wide range of conservation needs, using a flexible Bayesian
  hierarchical generalized additive model}.
\newblock {\em The Condor}, 123(1):duaa065.

\bibitem[Sohl et~al., 2016]{sohl2016modeled}
Sohl, T., Reker, R., Bouchard, M., Sayler, K., Dornbierer, J., Wika, S.,
  Quenzer, R., and Friesz, A. (2016).
\newblock Modeled historical land use and land cover for the conterminous
  united states.
\newblock {\em Journal of Land Use Science}, 11(4):476--499.

\bibitem[Spake et~al., 2023]{spake2023understanding}
Spake, R., Bowler, D.~E., Callaghan, C.~T., Blowes, S.~A., Doncaster, C.~P.,
  Antao, L.~H., Nakagawa, S., McElreath, R., and Chase, J.~M. (2023).
\newblock Understanding ‘it depends’ in ecology: a guide to hypothesising,
  visualising and interpreting statistical interactions.
\newblock {\em Biological Reviews}.

\bibitem[Stanton et~al., 2018]{stanton2018analysis}
Stanton, R., Morrissey, C.~A., and Clark, R.~G. (2018).
\newblock Analysis of trends and agricultural drivers of farmland bird declines
  in north america: A review.
\newblock {\em Agriculture, Ecosystems \& Environment}, 254:244--254.

\bibitem[Stein, 2014]{stein2014limitations}
Stein, M.~L. (2014).
\newblock Limitations on low rank approximations for covariance matrices of
  spatial data.
\newblock {\em Spatial Statistics}, 8:1--19.

\bibitem[Sultaire et~al., 2022]{sultaire2022spatial}
Sultaire, S.~M., Humphreys, J.~M., Zuckerberg, B., Pauli, J.~N., and Roloff,
  G.~J. (2022).
\newblock Spatial variation in bioclimatic relationships for a snow-adapted
  species along a discontinuous southern range boundary.
\newblock {\em Journal of Biogeography}, 49(1):66--78.

\bibitem[Thorson, 2019]{thorson2019guidance}
Thorson, J.~T. (2019).
\newblock {Guidance for decisions using the Vector Autoregressive
  Spatio-Temporal (VAST) package in stock, ecosystem, habitat and climate
  assessments}.
\newblock {\em Fisheries Research}, 210:143--161.

\bibitem[Thorson et~al., 2023]{thorson2023spatially}
Thorson, J.~T., Barnes, C.~L., Friedman, S.~T., Morano, J.~L., and Siple, M.~C.
  (2023).
\newblock Spatially varying coefficients can improve parsimony and descriptive
  power for species distribution models.
\newblock {\em Ecography}, page e06510.

\bibitem[Tyre et~al., 2003]{tyre2003improving}
Tyre, A.~J., Tenhumberg, B., Field, S.~A., Niejalke, D., Parris, K., and
  Possingham, H.~P. (2003).
\newblock Improving precision and reducing bias in biological surveys:
  estimating false-negative error rates.
\newblock {\em Ecological Applications}, 13(6):1790--1801.

\bibitem[Valavi et~al., 2021]{valavi2021modelling}
Valavi, R., Elith, J., Lahoz-Monfort, J.~J., and Guillera-Arroita, G. (2021).
\newblock Modelling species presence-only data with random forests.
\newblock {\em Ecography}, 44(12):1731--1742.

\bibitem[von Hirschheydt et~al., 2023]{von2023mixed}
von Hirschheydt, G., Stofer, S., and K{\'e}ry, M. (2023).
\newblock {``Mixed'' occupancy designs: When do additional single-visit data
  improve the inferences from standard multi-visit models?}
\newblock {\em Basic and Applied Ecology}, 67:61--69.

\bibitem[Watanabe, 2010]{watanabe2010}
Watanabe, S. (2010).
\newblock Asymptotic equivalence of bayes cross validation and widely
  applicable information criterion in singular learning theory.
\newblock {\em Journal of Machine Learning Research}, 11(12).

\bibitem[Wood, 2006]{wood2006generalized}
Wood, S.~N. (2006).
\newblock {\em {Generalized additive models: an introduction with R}}.
\newblock Chapman and Hall/CRC.

\bibitem[Zipkin et~al., 2012]{zipkin2012evaluating}
Zipkin, E.~F., Grant, E. H.~C., and Fagan, W.~F. (2012).
\newblock Evaluating the predictive abilities of community occupancy models
  using auc while accounting for imperfect detection.
\newblock {\em Ecological Applications}, 22(7):1962--1972.

\end{thebibliography}

\newpage

\section*{Figures}

\newpage

\newpage

\newpage

\newpage

\end{document}